\documentclass[onecolumn,showpacs,preprintnumbers,amsmath,amssymb]{revtex4}

\usepackage{graphicx}
\usepackage{dcolumn}
\usepackage{bm}

\begin{document}

\title{UNQUENCHING THE QUARK MODEL}

\author{E.~SANTOPINTO}

\affiliation{I.N.F.N. and Dipartimento di Fisica,\\ via Dodecaneso 33,
Genova, I-16146, ITALY\\ 
E-mail: santopinto@ge.infn.it}

\author{R.~BIJKER}
\affiliation{
Departamento de Estructura de la Materia, Instituto de Ciencias Nucleares, \\  
Universidad Nacional Aut\'onoma de M\'exico, AP 70-543, 04510 M\'exico DF, MEXICO\\
E-mail: bijker@nucleares.unam.mx} 
\begin{abstract}
We present the formalism for a new generation of unquenched quark models in which quark-antiquark 
pair effects are taken into account in an explicit form via a microscopic QCD-inspired quark-antiquark 
creation mechanism. No truncation in the sum over all the big tower of states is necessary since these  states are automatically generated by means of powerful group-theoretical techniques.
An important check on the formalism and the numerical results is provided by the closure limit. 
As an application, the effect of quark-antiquark pairs on the strange content of the 
proton spin, $\Delta s$, is discussed. The contributions 
of the up and down quarks, $\Delta u$ and $\Delta d$, are also calculated. This has become possible 
after solving the difficult problem of permutational symmetry related to quark 
rearrangements. Finally, we present some preliminary results in the closure limit 
as well as an outlook for future applications.
\end{abstract} 

\maketitle
\section{Introduction}
Since the QCD equations in the non-perturbative region are not solvable, 
many  effective models of the hadrons have been developed, 
in particular many versions of the Constituent Quark Model (CQM). 
These are able to give good results for the static properties of the 
hadrons (like the spectrum and the magnetic moments), while they 
all fail to reproduce the dynamic ones, like the electromagnetic 
transition form factors at low $Q^2$ values. 
This seems to be a 
problem of degrees of freedom, since the region of low $Q^2$ means high 
distance, in which the creation of quark-antiquark-pair degrees has higher 
probability.  

A priori, one would expect pair creation to be (after switching on a 
pair-creation mechanism) so enhanced in probability that a 
valence quark model would fail dramatically. 
On the other hand, it is known from the phenomenology that pair 
creation is suppressed, at least at the zeroth order in the static observables, 
since the observed spectrum is dominated by the valence quarks $qqq$ for the 
baryons and by the valence $q \bar q$ for the mesons. 
Moreover, the resonances do not have decay widths of the order of the same resonance masses.
Many years ago, Isgur tackled the intriguing and important question of  
why the CQMs work so well at an intermediate regime.
The aim of this paper is to continue from where he left off and to develop the formalism for the light baryons with the linked problem of the permutation symmetry.
\section{Degrees of freedom}
Constituent Quark Models based on the effective degrees of freedom of 
three constituent quarks have been proposed in several versions: the Isgur-Karl model 
\cite{IK}, the Capstick-Isgur model \cite{capstick}, the algebraic $U(7)$ model \cite{bil}, 
the hypercentral model \cite{pl}, the chiral boson exchange model 
\cite{olof,ple} and the Bonn instanton model \cite{bn}. 
While these models display important and peculiar differences, they all have main features in common:
they are all based on the effective degrees of freedom of three constituent quarks and 
on the $SU(6)$ spin-flavor symmetry; they also contain a long-range
linear confining potential and an $SU(6)$-breaking term, even though the form and the advocated origin of
this last term may be different, as in the one-gluon-exchange-inspired hyperfine interaction or
the Goldston Boson Exchange $SU(6)$-breaking interaction or the instanton-induced-breaking term.

The photocouplings calculated by means of different constituent quark models (see for example 
Table~2 of Ref. \cite{aie}, and references therein) have the same overall behaviour,
having the same $SU(6)$ structure in common, but in many cases they show a lack of strength.

\begin{figure}[ht]
\vspace{-0.4cm}
\centering
\includegraphics[width=6.5cm]{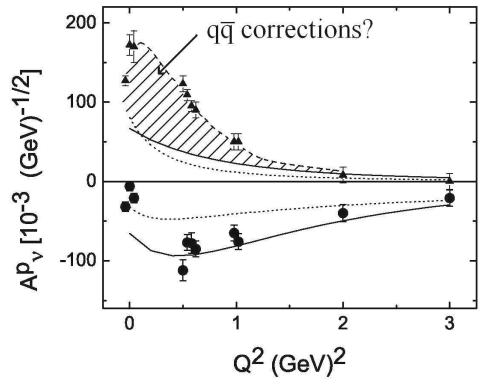} 
\caption[]{\small Transition form factors for the $D_{13}(1520)$ resonance. Experimental data are compared with theoretical predictions from some CQMs, here the collective $U(7)$ \cite{bil2}, 
dotted line, and the hypercentral \cite{aie2}, unbroken line. The missing strength at low $Q^2$ is interpreted as due to $q\bar{q}$ pair effects.}
\label{d13}
\vspace{-0.2cm}
\end{figure}

For each of those models for which calculations of the electromagnetic transition 
form factors are available, there is clearly the problem of missing strength at low
$Q^2$, as is well illustrated in Fig.~\ref{d13}, which reports the transverse 
electromagnetic transition form factors for the $D_{13}(1520)$ resonance. 
The experimental data are compared with the predictions of different CQMs. 
The common problem of missing strength at low $Q^2$ can be ascribed to the lack 
of quark-antiquark effects, which are probably more important in the 
outer region of the nucleon.
After understanding that the main problems of the CQMs are all linked to a problem of missing-degrees of 
freedom, and after identifing the key degrees of freedom as $q\bar{q}$ pairs, we are faced with two possibilities: the phenomenological parametrization  or the microscopic 
explicit description. We have focused on the letter and tried to find a QCD-inspired 
pair-creation mechanism that at the same time should be encoded in such a way 
not as not to destroy the good CQM results. Moreover, it is important to find symmetry constraints 
in order to help to keep all the difficult calculations under control. 
In this respect, the closure limit, is very helpful, as will be explained in the following sections. 
\section{Flux-tube breaking model}

In the flux-tube model for hadrons, the quark potential model arises from an adiabatic 
approximation to the gluonic degrees of freedom embodied in the flux tube \cite{flux}. 
The impact of $q \bar{q}$ pairs in meson spectroscopy has been studied in an elementary 
flux-tube breaking model \cite{mesons} in which the quark pair creation occurs with $^{3}P_0$ 
quantum numbers. Subsequently, it has been shown by Geiger and Isgur \cite{OZI} that a 
"miraculous" set of cancellations between apparently uncorrelated sets of mesons occurs 
in such a way to compensate each 
other and not to destroy the good CQM results for the mesons, and in particular in such a 
way that (i) the OZI hierarchy is preserved and (ii) there is a near immunity of the 
long-range confining potential.
The change in the linear potential due to pairs bubbling in the string can be reabsorbed 
in a new strength of the linear potential, {\em i.e.} in a new string tension; thus the 
net effect of the mass shifts from pair creation is smaller than the naive expectation of 
the order of the strong decay widths.
The important point is that it is necessary to sum over very large towers of intermediate 
states to see that the spectrum of the mesons is only weakly perturbed, after unquenching 
and renormalizing. 
No simple truncation of the set of meson loops can reproduce such results.
\begin{figure}[ht]
\vspace*{-4.2cm}
\centering
\includegraphics[width=9cm]{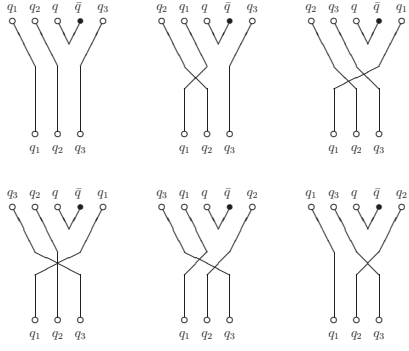} 
\vspace*{-4cm}
\caption[]{\small Quark line diagrams for $A \rightarrow B C$}
\label{diagrams}
\end{figure}

This formalism set up for mesons could not be immediately 
extended to  baryons, since there were extra problems, mainly linked with 
the permutation symmetry between identical quarks. 
In \cite{baryons}, Geiger and Isgur investigated the importance of $s \bar{s}$ 
loops in the proton by taking into account the contribution of the six different 
diagrams of Fig.~\ref{diagrams} (in Fig.~3 of \cite{baryons} only four diagrams are 
considered) and by using harmonic oscillator wave functions for the baryons 
and mesons. 

In the present manuscript, we discuss some generalizations of this formalism for 
baryons. These new extensions make it possible to study the quark-antiquark contributions 
\begin{itemize}
\item for any initial baryon resonance
\item for any flavor of the quark-antiquark pair
\item for any model of baryons and mesons, as long as their wave functions are 
expressed on the basis of the harmonic oscillator.
\end{itemize}
The problem of the permutation symmetry between identical quarks has been solved by 
means of group-theoretical techniques. In this way, we can evaluate the contribution 
of all the diagrams for any initial baryon $q_1 q_2 q_3$ 
(ground state or resonance) and for any flavor of the $q \bar{q}$ and  not just
for the $s\bar{s}$.

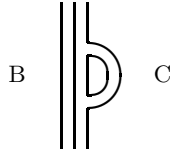
\begin{figure}[ht]
\centering
\setlength{\unitlength}{0.5pt}
\begin{picture}(120,150)(50,0)
\thicklines
\put(100, 20) {\line(0,1){110}}
\put(110, 20) {\line(0,1){110}}
\put(120, 20) {\line(0,1){30}}
\put(120, 60) {\line(0,1){30}}
\put(120,100) {\line(0,1){30}}
\put(120, 75) {\oval(30,30)[br]}
\put(120, 75) {\oval(30,30)[tr]}
\put(120, 75) {\oval(50,50)[br]}
\put(120, 75) {\oval(50,50)[tr]}
\put( 60, 70) {B}
\put(170, 70) {C}
\end{picture}
\caption[]{\small One-loop diagram for the meson correction at the quark level}
\label{loop}
\end{figure}

The pair-creation mechanism is inserted at the level of quarks and the one-loop 
diagram should be calculated, see Fig.~\ref{loop}, by summing over all the possible 
intermediate states. This is the crucial step, after solving all the problems linked 
with this infinite sum in the case of the baryons. 

Isgur has shown that as a first approximation we can think of the baryons as the sum of the
zeroth order three-quark Fock component plus the sum over all the possibile intermediate 
states due to the creation of quark-antiquark pairs.  
Here we adopt a pair-creation model to incorporate the effects of 
$q \bar{q}$ pairs. Up to leading order in pair creation 
the baryon wave function is given by \cite{baryons}
\begin{eqnarray} 
\mid \psi_A \rangle \;=\; {\cal N} \left[ \mid A \rangle 
+ \sum_{BC l J} \int d \vec{k} \, \mid BC \vec{k} \, l J \rangle \, 
\frac{ \langle BC \vec{k} \, l J \mid h_{q \bar{q}}^{\dagger} \mid A \rangle } 
{M_A - E_B - E_C} \right] ~,
\end{eqnarray}
where $h_{q \bar{q}}^{\dagger}$ is the quark-antiquark pair-creation operator as in \cite{baryons} , 
$A$ is the initial baryon, $B$ ($C$) denotes the intermediate baryon (meson), 
$\vec{k}$ and $l$ the relative radial momentum and orbital angular 
momentum of $B$ and $C$, and $J$ is the total angular momentum 
$\vec{J} = \vec{J}_{BC} + \vec{l} = \vec{J}_B + \vec{J}_C + \vec{l}$. 
In general, matrix elements of observables $\hat{\cal O}$ are given by
\begin{eqnarray}
{\cal O} \;=\; \langle \psi_A \mid \hat{\cal O} \mid \psi_A \rangle 
\;=\; {\cal O}_{\rm valence} + {\cal O}_{\rm sea} 
\end{eqnarray}
where the first term denotes the contribution from the three valence quarks  
\begin{eqnarray}
{\cal O}_{\rm valence} &=& {\cal N}^2 \langle A \mid \hat{\cal O} \mid A \rangle 
\end{eqnarray}
and the second term the contribution from the quark-antiquark pairs
\begin{eqnarray}
{\cal O}_{\rm sea} &=& {\cal N}^2 \sum_{BC l J} \int d \vec{k} \,
\sum_{B' C' l' J'} \int d \vec{k}^{\, \prime} \,
\frac{ \langle A \mid h_{q \bar{q}} \mid B' C' \vec{k}^{\, \prime} \, l' J' \rangle } 
{M_A - E_{B'} - E_{C'}} 
\nonumber\\
&& \hspace{2cm} \langle B' C' \vec{k}^{\, \prime} \, l' J' \mid {\cal O}  
\mid B C \vec{k} \, l J \rangle \, 
\frac{ \langle B C \vec{k} \, l J \mid h_{q \bar{q}}^{\dagger} \mid A \rangle } 
{M_A - E_B - E_C} ~.
\label{me}
\end{eqnarray}
The sum is over a complete set of intermediate states, 
rather than just a few low-lying states. Not only does this have a significant 
impact on the numerical result, but it is necessary for consistency with the 
OZI-rule and the success of the CQMs in spectroscopy
This is possible since the intermediate states are generated automatically by means of group-theory 
techniques. It has been implemented in a computer code which is able to construct  
states with the correct pemutational symmetry for any model and up to any shell. 
Therefore, the sum over intermediate states can be perfomed up to saturation and not 
just for the first few shells as in \cite{baryons}. 

\section{Closure limit}
In this section, we discuss the closure limit which arises 
when the energy denominators do not depend strongly on the quantum numbers of the intermediate states in Eq.~(\ref{me}). 
In this case, the sum over the complete set of intermediate states can 
be solved by closure and the contribution of the quark-antiquark pairs to 
the matrix element reduces to 
\begin{eqnarray}
{\cal O}_{\rm sea} &\propto& \langle A \mid h_{q \bar{q}} \, {\cal O} \,   
h_{q \bar{q}}^{\dagger} \mid A \rangle ~.
\end{eqnarray}
This means for example that, at least  qualitatively, we can already say that the 
strange content of the proton should be very small, since within 
the closure limit it comes out to be exactly zero. The closure limit is very important for many reasons 
that are closely interconnected. It is a very good stringent test that a 
numerical calculation should fulfill. Moreover, and even more important, it also explains, 
when it comes out equal to zero, the success of the CQMs.
The corrections due to the $q\bar{q}$ pairs are zero in the closure limit for many observables, 
while the sums over big tower of states are constrained by the closure limit in such a 
way that very differt meson and baryon states compensate one another.

\section{Preliminary results} 
\begin{figure}[ht]
\centering
\setlength{\unitlength}{0.5pt}
\begin{picture}(440,200)(20,80)
\thicklines
\put(100,250) {\line(1,0){300}}
\put(150,200) {\line(1,0){200}}
\put(200,150) {\line(1,0){100}}

\put(150,200) {\line(1,1){ 50}}
\put(200,150) {\line(1,1){100}}
\put(250,100) {\line(1,1){150}}

\put(250,100) {\line(-1,1){150}}
\put(300,150) {\line(-1,1){100}}
\put(350,200) {\line(-1,1){ 50}}

\multiput(100,250)(100,0){4}{\circle*{10}}
\multiput(150,200)(100,0){3}{\circle*{10}}
\multiput(200,150)(100,0){2}{\circle*{10}}
\put(250,100){\circle*{10}}

\put( 85,265){$ddd$}
\put(185,265){$udd$}
\put(285,265){$uud$}
\put(385,265){$uuu$}
\put(135,215){$dds$}
\put(235,215){$uds$}
\put(335,215){$uus$}
\put(185,165){$dss$}
\put(285,165){$uss$}
\put(235,115){$sss$}

\put(0,250){$\Delta$}
\put(0,200){$\Sigma^{\ast}$}
\put(0,150){$\Xi^{\ast}$}
\put(0,100){$\Omega$}

\end{picture}
\caption[]{\small Ground state decuplet baryons}
\label{decuplet}
\end{figure}
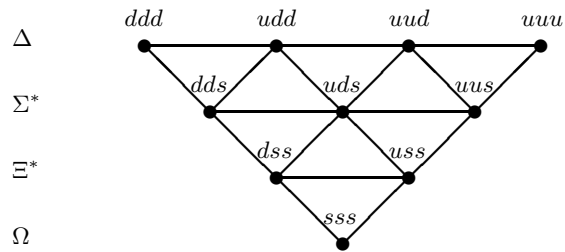

In this section, we discuss some preliminary results in the closure 
limit for the operator $\Delta q$ to determine the fraction of the baryon's 
spin carried by each one of the flavors $u$, $d$ and $s$
\begin{eqnarray}
\Delta q \;=\; 2 \langle S_z(q) + S_z(\bar{q}) \rangle ~.
\end{eqnarray}
In the closure limit the relative contribution of the quark flavors from the 
quark-antiquark pairs to the baryon spin is the same as that from the 
valence quarks
\begin{eqnarray}
\Delta u_{\rm sea} : \Delta d_{\rm sea} : \Delta s_{\rm sea} \;=\; 
\Delta u_{\rm valence} : \Delta d_{\rm valence} : \Delta s_{\rm valence} ~.
\end{eqnarray}
Table~\ref{baryonspin} shows the relative contributions of $\Delta u$, 
$\Delta d$ and $\Delta s$ to the spin of the ground state octet baryons  
in the closure limit. 
\begin{table}[ht]
\centering
\caption[]{\small Relative contributions of $\Delta u$, $\Delta d$ and $\Delta s$ 
to the spin of the ground state octet baryons with $^{2}8[56,0^+]_J$ 
in the closure limit}   
\label{baryonspin}
\begin{tabular}{crcrcr}
\hline
& $\Delta u$ &:& $\Delta d$ &:& $\Delta s$ \\
\hline
$p$ & 4 &:& $-1$ &:& 0 \\
$n$ & $-1$ &:& 4 &:& 0 \\
$\Sigma^+$ & 4 &:& 0 &:& $-1$ \\
$\Sigma^0$ & 2 &:& 2 &:& $-1$ \\
$\Sigma^+$ & 0 &:& 4 &:& $-1$ \\
$\Lambda$ & 0 &:& 0 &:& 3 \\
$\Xi^0$ & $-1$ &:& 0 &:& 4 \\
$\Xi^-$ & 0 &:& $-1$ &:& 4 \\
\hline
\end{tabular}
\end{table}
The closure limit, when combined with symmetries, imposes other strong limits, 
as for example can be seen for the ground state decuplet baryons in Fig.~\ref{decuplet}.
For the $\Delta$ resonances whose three-quark configuration does not contain strange quarks,  
the contribution of the $s \bar{s}$ pairs to the spin $\Delta s$ and the magnetic moment 
$\mu_s$ vanishes in the closure limit. The same holds for the contribution of $d \bar{d}$ 
pairs to the $\Delta^{++}$, $\Sigma^{\ast \, +}$, $\Xi^{\ast \, 0}$ and $\Omega^{-}$ resonances, 
and that of $u \bar{u}$ pairs to the $\Delta^{-}$, $\Sigma^{\ast \, -}$, $\Xi^{\ast}$$^{-}$ and $\Omega^{-}$ resonances. 
These are very stringent requirements since each one involves the sum over a complete set of 
 states of the intermediate baryon-meson system and provides a nontrivial test which 
involves the spin-flavor section, the permutation symmetry and the correct implementation of 
large towers of intermediate states. 

\section{Conclusions and outlook}
The formalism described in this paper will be systematically used in order to study the still open problems regarding light baryons, such as the sea quark content and the spin problem of the proton \cite{new} and the $q\bar{q}$ effects in the electromagnetic elastic 
and transition form factors \cite{new1}.  

\section*{Acknowledgments}
This work was supported in part by the I.N.F.N. and in part by a grant from CONACyT, Mexico.

\end{document}